\pdfoutput=1
\documentclass[letterpaper]{article}
\usepackage{aaai17}
\usepackage{latexsym}
\usepackage{color}
\usepackage[dvipsnames,table]{xcolor}
\usepackage{graphicx}
\usepackage{times}
\usepackage{helvet}
\usepackage{courier}
\usepackage{subcaption}
\usepackage{booktabs}
\usepackage{float}
\usepackage[T1]{fontenc}
\usepackage[utf8]{inputenc}
\usepackage[russian,english]{babel}
\usepackage{ctable,multirow}
\usepackage{comment}
\floatstyle{boxed} 
\restylefloat{figure}

\newcommand{\citeAuthorYear}[1]{\citeauthor{#1} (\citeyear{#1})}

\begin{document}

\title{Measuring, Predicting and Visualizing Short-Term Change in Word Representation and Usage in VKontakte Social Network} 

\author{
Ian Stewart\\
Georgia Institute of Technology\\
801 Atlantic Dr NW\\
 Atlanta, GA 30332 \\
\And
Dustin Arendt, Eric Bell and Svitlana Volkova\\
Pacific Northwest National Laboratory\\
902 Battelle Blvd\\
 Richland, WA 99354\\
}

\maketitle

\begin{abstract}
Language in social media is extremely dynamic: new words emerge, trend and disappear, while the meaning of existing words can fluctuate over time. Such dynamics are especially notable during a period of crisis. This work addresses several important tasks of measuring, visualizing and predicting short term text representation shift, i.e. the change in a word's contextual semantics, and contrasting such shift with surface level word dynamics, or concept drift, observed in social media streams. Unlike previous approaches on learning text representations in text, we study the relationship between short-term concept drift and representation shift on a large social media corpus -- VKontakte posts in Russian collected during the Russia-Ukraine crisis in 2014 -- 2015. Our novel contributions include quantitative and qualitative approaches to (1) measure short-term representation shift and contrast it with surface level concept drift; (2) build predictive models to forecast short-term shifts in meaning from previous meaning as well as from concept drift; and (3) visualize short-term representation shift for example keywords to demonstrate the practical use of our approach to discover and track meaning of newly emerging terms in social media. We show that short-term representation shift can be accurately predicted up to several weeks in advance. Our unique approach to modeling and visualizing word representation shifts in social media can be used to explore and characterize specific aspects of the streaming corpus during crisis events and potentially improve other downstream classification tasks including real-time event detection.

 \end{abstract}

\section{Introduction}

Social media have been widely studied as sensors of human behavior to track unusual or novel activities in real time all over the globe~\cite{Alsaedi:16,Asur:10predicting}. Much analysis of social media language focuses on surface-level features and patterns, like word frequency, to improve real-time event detection and tracking e.g., during crisis events~\cite{Bruno:11}, elections~\cite{Lampos:13}, and natural disasters~\cite{Crooks:13earthquake}. These surface features provide a shallow signal into human behavior~\cite{Eisenstein:14} but miss some of the more subtle variations. Tracking emerging words only based on their virality and frequency trends~\cite{Mathioudakis:10,Weng:13}, or using dynamic topic models~\cite{Blei:06} would miss the change in word meaning for existing words, or the meaning of newly emerging terms in social media. 
For example, during the Russian-Ukrainian crisis in 2014 -- 2015 the word {\it ukrop}, meaning {\it dill}, changed its meaning to {\it Ukrainian patriot} and developed a more negative connotation over time. Recent work has effectively tracked word meaning over time at scale, but it often examines long-term meaning shift within formal written text such as the Google Books corpus~\cite{Gulordava:11}, rather than short-term shift within more informal contexts~\cite{Kulkarni:15}. 

The goal of this work is to analyze short-term meaning change and frequency dynamics jointly over time using a corpus of VKontakte posts collected during the Russia-Ukraine crisis, which has been noted as a source of political instability~\cite{Duvanova:15,Volkova:16Dec} and thus linguistic unpredictability. We develop an approach for predicting and tracking short-term shifts in word meaning or {\it representation}, utilizing a word's previous meaning as well as the word's frequency in social media streams. 

We demonstrate that combining both shallow and deep metrics of language variation can maximize the signal from social media and help uncover more subtle changes in language use. 
Our unique approach to modeling and visualizing word representation shift can explain how a word's context shifts over time rather then just tracking its tendency to trend: e.g., the word {\it fire} might initially be semantically associated with {\it water} but over time grow more associated with {\it bombing} and {\it attack}. Moreover, accurately predicting word representation shift can help to identify dynamic words whose meaning is on the verge of changing. Tracking these subtle changes could in turn benefit downstream applications like event detection, which often relies on the frequency rather than the meaning of words~\cite{Alsaedi:16,Corney:14}. 

As an application area for this work, we focus on monitoring word meaning change over time during crises.\footnote{Recent crises include Arab Spring in Egypt and Tunisia in 2010 and 2011, Russian elections and Bolotnaya protests in 2011, Russian takeover of Crimea in 2014, conflicts in Iran and Syria.} As crises are increasingly discussed over Twitter and other social media, the ability to automatically measure word changes relevant to these events is extremely important. 
With events that unfold quickly, analysts can become quickly overwhelmed with the volume of data available and may miss subtle changes in common words such as {\it ukrop}, that can take on unexpected meanings during a crisis.
Systems that can assist analysts during these events could bring us much closer to understanding issues as they happen on the ground.



\begin{figure}[t!]
\centering
\includegraphics[width=0.35\textwidth]{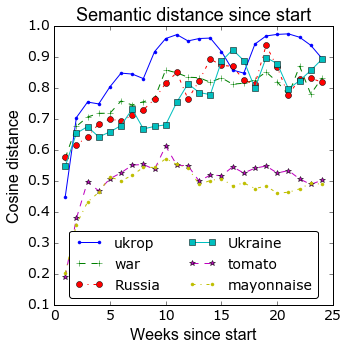}
\caption{Motivation example: representation shift measured using cosine distance between each word's current representation and its original representation over time.}
\label{fig:motivation_shift}
\end{figure}

In our study, we distinguish {\it concept drift}~\cite{Gama:14}, a change in a word's frequency over time from {\it representation shift}, a change in its semantic representation estimated from context. We quantify representation shift as the distance between time-aware word embeddings~\cite{Mikolov:13}, which are distributed representations of words in low-dimensional vector space at adjacent timesteps. Work in distributional semantics has proposed that words with similar contexts also exhibit similar meaning~\cite{Bengio:03,Harris:54}. Thus, semantically similar words should have similar embeddings or vector representations; e.g., vectors for the foods {\it dill} and {\it tomato} should be close together. 
Tracking word embeddings over time, we propose that representation shift can reveal word dynamics that would have gone unseen
and can be detected on the scale of weeks rather than years~\cite{Hamilton:16Diachronic}. 

To address the phenomenon of representation shift, our study makes the following novel contributions:
\begin{itemize}
\item We relate computational measures of concept drift and representation shift (i.e. meaning change) to track language dynamics in social media.
\item We find common trajectories in concept drift and representation shift by clustering words by their dynamics into three categories -- increasing, decreasing and constant.
\item We develop predictive models and show that short-term representation shift can be effectively inferred from prior shift and concept drift.
\item We propose novel visual representations to track the development of new words in social media as a useful tool for analysts to explore unexpected language change and text connotations.
\end{itemize}

\subsection{Motivation}

To motivate the study, we present in Figure \ref{fig:motivation_shift} an example of representation shift in a set of keywords related to food and to the Russian-Ukrainian crisis, drawn from VKontakte social network data (see ``Data'' section). We use cosine distance from the first timestep's embedding as a proxy for representation shift. First, we note the split between the relatively stable food words (i.e. minimal cosine distance) and the more dynamic conflict words that become increasingly distant from their original representation. Moreover, we see that some of the upper words such as {\it ukrop} exhibit especially dynamic behavior, alternatively growing farther and closer to its original meaning from weeks 11 to 17. This shift was likely the result of a split in meaning as {\it ukrop} was used less as its literal meaning {\it dill} and more as a slang pejorative for {\it Ukrainian patriot}.

\begin{figure}[t!]
	\centering
	\includegraphics[width=0.375\textwidth]{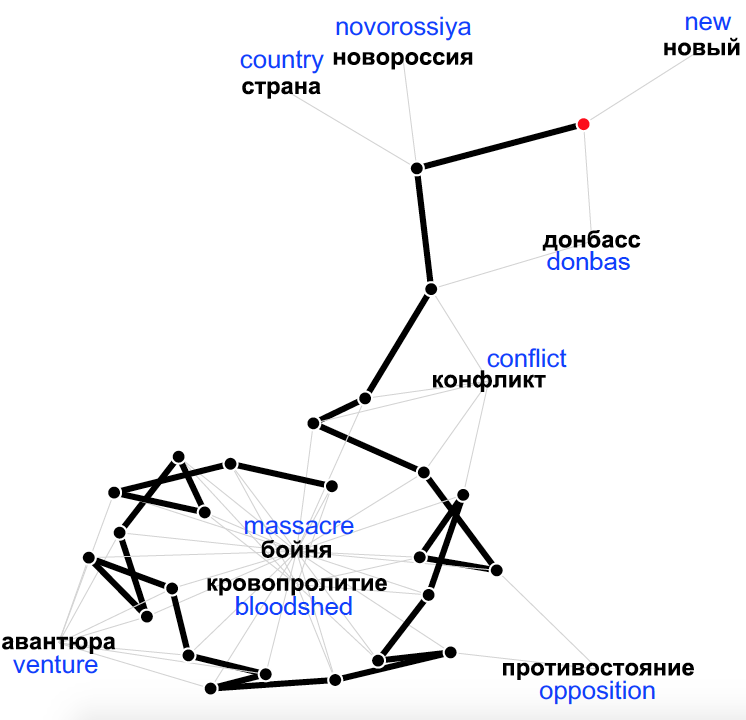}
	\caption{Semantic trajectory of the word {\it war} over time, projected in 2D using principal component analysis, with two most similar words at each timestamp.}
	\label{fig:motivation_trajectory}
	\vspace{-0.2cm}
\end{figure}

We draw further motivation from Figure \ref{fig:motivation_trajectory}, which illustrates the trajectory of representation shift 
in the word {\it war} starting from the upper-right red dot and progressing toward the end of the line in the lower-left. Projected into 2 dimensions, the embedding of {\it war} begins close to the embeddings of situation-specific words such as {\it Donbas} (location of conflict) and cycles toward more violent words such as {\it massacre}. This context shift appears to be a kind of semantic narrowing~\cite{Sagi:09} toward more negative or pejorative words that is captured in the decreasing shift distances in the later timesteps. Similar to the  changes in Figure \ref{fig:motivation_shift}, the narrowing in {\it war} may be the result of increased public interest in the topic of conflict, first causing location-specific discussion and later leading to polarization of the word.
\section{Background}
In studying representation shift, we draw on prior work on concept drift which has approached the problem of predicting word frequency changes with a variety of methods. For example,~\citeAuthorYear{Costa:14} proposed models to predict behavior in Twitter hashtag frequency across predefined drift categories such as gradual and incremental. Studies in concept drift have also worked to develop adaptive models~\cite{Gama:14}, such as~\citeAuthorYear{Magdy:14} who develop a classification technique that adapts to dynamic topics by adopting new keywords into an initial query set, according to their estimated relevance. Our work suggests that research in concept drift can benefit from incorporating not just word frequency but also word context.
We also build on prior attempts to predict noisy time series data~\cite{Oancea:14}, such as fluctuations in stock prices, through machine learning methods. 

In addition, our work leverages techniques from distributional semantics to approximate the change in word representation over time. Studies in computational semantics focus primarily on the use of language in a specific timespan or domain~\cite{Sagi:09} rather than examining the variation in word meanings over time.~\citeAuthorYear{Kim:14} propose a novel approach by measuring change in English word semantics across the 20th century by comparing each word's initial meaning (measured by its embedding) with its meanings in later years. Further studies adopt this methodology to propose laws of semantic change relating concept drift to representation shift~\cite{Hamilton:16Diachronic} as well as to separate insignificant from significant linguistic change across domains~\cite{Kulkarni:15}. 

Our work builds on prior studies by first tracking semantic change within a non-English language and in the noisy domain of social media rather than non-social corpora like Google Books~\cite{Hamilton:16Diachronic,Kim:14}. We also seek to explain previously unknown semantic change instead of previously known changes (e.g., {\it gay})~\cite{Hamilton:16Cultural}. In addition, we look to highlight more subtle, short-term changes in connotation, such as the added pejorative connotation of {\it ukrop } ({\it dill}) as it became a negative descriptor for {\it Ukrainian patriot}. Because of the socially-situated nature of language~\cite{Eisenstein:14}, we believe that these fine-grained changes are pervasive in social media data and deserve more exploration for the sake of applications like event detection. Lastly, our study is among the first to build predictive models to forecast representation shift rather than characterizing it~\cite{mitra2015automatic}.

\section{Data}
We rely on public data from the VKontakte (VK) social network, a popular website in Russia and Ukraine similar to Facebook~\cite{Volkova:16Dec}. The data were collected
over a period of 25 weeks between September 2014 and March 2015, and comprise over 600,000 posts produced by 50,000 users,  with 51\% of posts from Russia and 49\% from Ukraine. VK does not have any restrictions on post length, thus VK messages are usually longer than tweets: the average post length is 167 words and 1107 characters. VK posts do not contain any user mentions or hashtags, but they contain URLs and attributes indicative of informal language including abbreviations and slang. 

The VKontakte data provides an ideal testbed for our study of representation shift because it was collected during a volatile period in Russia and Ukraine that led to sudden language change, such as the adoption of new meaning for words like {\it ukrop}. Nonetheless, our methods can apply to any active social media platform that uses primarily text data for user interaction, such as Twitter or Facebook.

\paragraph{Data Pre-processing} 
Following standard practices, we first stem all words in the data using the Russian morphology package PyMorph\footnote{https://pymorphy2.readthedocs.io/en/latest/} and lower-case the words to match social media's irregular capitalization practices. We then collect all unigrams with a post frequency of 5 and above to avoid the long tail of misspellings and irrelevant words, leaving us with a vocabulary $V$ of about 60,000 words. We also remove Russian and Ukrainian stop-words from frequency counts but do not remove them from the word vector vocabulary in order to preserve the context afforded by stop-words. When necessary during analysis, we consulted various online dictionaries for translations of slang words that were otherwise difficult to parse. For instance, at the time of writing Google Translate only provided the literal translation of {\it ukrop} as {\it dill}.

\section{Methods}

\subsection{Measuring Word Usage and Meaning}
\subsubsection{Word Usage Dynamics}
Frequency-based methods can capture linguistic shift because changes in word frequency often correspond to words gaining or losing senses~\cite{Hamilton:16Diachronic}. Thus, we first extract weekly frequency-based statistics, word frequency $f$ and tf-idf score $\chi$, to capture changes in word usage in social media without over-representing words with consistently high frequency. We restrict our study to these basic statistics 
in order to test intuitive assumptions about representation shift on social media. We define the statistics for each word $w$ over all posts $P_{t}$ at week $t$ as follows:
\begin{equation} f_{w, t} = \frac{count(w,t)}{\sum_{w \in V} count(w, t)}.\end{equation}

\begin{equation}\chi_{w, t} = log(count(w,t)) \times log\frac{|P_{t}|}{|{p \in P_{t} : w \in p}|}.\end{equation}

Lastly, the usage statistics for each word $w$ are concatenated chronologically to form time series $\tau_{f}(w)$ and $\tau_{\chi}(w)$, which represent our measure of concept drift.

\subsubsection{Temporal Embeddings} 
We learn temporal representations of social media words by relying on recent advances in distributional semantics. We applied word2vec models~\cite{Mikolov:13} implemented in gensim\footnote{https://pypi.python.org/pypi/gensim}~\cite{Rehurek:10} that have been successfully used for a variety of applications related to tracking semantic change~\cite{Hamilton:16Cultural,Hamilton:16Diachronic,Kim:14,Kulkarni:15}. As an alternative to embeddings, it is possible to represent a word's meaning with more complicated models such as the set of its different senses~\cite{Frermann:16}, or as the distribution over different semantic ``spaces'' and topics~\cite{Blei:06,Kenter:15}. We chose word embeddings for their simple mathematical representation and well-understood applications for encoding meaning~\cite{Gulordava:11}. 

To begin, we initialize a model with vocabulary $V$, then train the model with tokenized posts for each timestep (i.e. week), using as a baseline the embeddings trained at the previous timestep. This guarantees that the dimensionality remains consistent across weeks and allows us to reliably track a word's representation shift through time. Formally, we train each timestep's embeddings until the model reaches the following threshold from epoch $\epsilon$ to $\epsilon-1$:
$$\rho = \frac{1}{|V|} \sum_{w \in V} arccos \frac{e_{ t,\epsilon} (w) \cdot e_{t,\epsilon-1} (w) } {|| e_{t,\epsilon} (w) || \cdot || e_{t,\epsilon-1} (w) ||}, $$

where $e_{t,\epsilon} (w)$ is the embedding vector for word $w$ at week $t$ at epoch $\epsilon$, and we set $\rho=0.0001$, the default setting. This training procedure allows the embeddings to encode the semantic change at timestep $t$ without overwriting the dimensionality of timestep $t-1$~\cite{Kim:14}. Thus, for each word in the vocabulary $V$, at each week $t$ between 1 and 25, we generate an embedding vector to yield a time series $\tau_e(w)$ as shown below:
\begin{equation}
\tau_e(w) =  e_{t_0} (w), e_{t_1} (w)  \dots  e_{T} (w).
\end{equation}

To build the embeddings, we chose a dimensionality of 30 for a fairly coarse-grained representation, avoiding data sparsity. We used standard training hyperparameters (e.g., window size 5), following prior experiments in building word vectors from social media data~\cite{Kulkarni:15}.


\subsubsection{Differencing Statistics}
\begin{figure}[t!]
	\centering
	\vspace{-0.1cm}
	\includegraphics[width=0.5\textwidth]{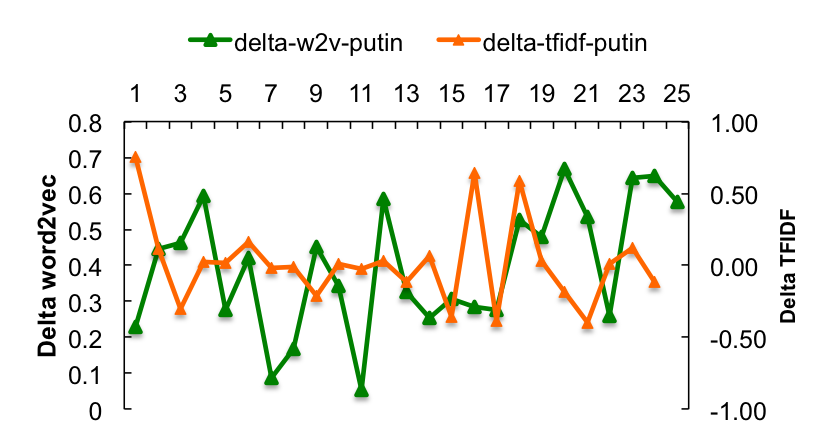}
	\caption{Concept drift $\Delta \tau_\chi(w)$ and representation shift $\Delta \tau_e(w)$ dynamics for the word {\it Putin}.}
	\label{fig:delta_w2v_tfidf}
	\vspace{-0.2cm}
\end{figure}

Our tests require us to compare representation shift with concept drift, and we therefore compute the first-order differences for all statistics ({\it e.g.,} word frequency at time $t$ and $t-1$). Formally, for each statistic $s$ in $\tau_f(w), \tau_\chi (w)$ and $\tau_e(w)$ over all words in vocabulary of size $N$, over the course of $T$ timesteps, we calculate a difference vector:
\begin{equation}
\Delta \tau_s(w) = \Delta s_{t_0,t_1} (w)  \dots \Delta  s_{T-1,T} (w),
\end{equation}

in which we calculate $\Delta\tau_f(w)$ and $\Delta\tau_\chi(w)$ with subtraction and $\Delta\tau_e(w)$ with cosine distance. Using the $\Delta$ rather than raw values for all statistics allows us to directly compare concept drift and representation shift in terms of their behavior between adjacent timesteps. We display an example of these statistics in Figure \ref{fig:delta_w2v_tfidf}, with the $\Delta\tau_\chi (w)$ (``Delta TFIDF'') and $\Delta\tau_e(w)$ (``Delta word2vec'') series for the sample word {\it Putin}. We see that the two types of change are not identical but do show a similar degree of ``spiky'' periods, although the $\Delta\tau_e(w)$ has a higher relative variance. Interestingly, the maximum $\Delta\tau_e(w)$ value occurs at timestep 20, following two spikes in $\Delta\tau_\chi(w)$. 


\begin{figure}[t!]
	\centering
	\vspace{-0.1cm}
	\includegraphics[width=0.41\textwidth]{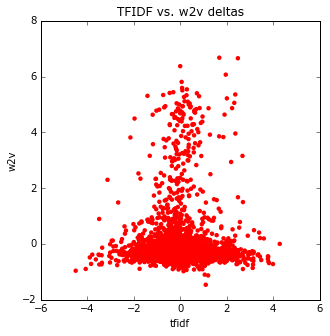}
	\caption{The distribution of representation shift $\Delta \tau_e(w)$ and concept drift in $\Delta \tau_\chi(w)$ dynamics.}
	\label{fig:tfidf_w2v_distribution}
	\vspace{-0.2cm}
\end{figure}

To provide a more general picture of the data, Figure \ref{fig:tfidf_w2v_distribution} presents a distribution of $\Delta \tau_\chi(w)$ versus $\Delta \tau_e(w)$ for a sample of the vocabulary and timesteps. We see that the tf-idf dynamics tend toward a normal distribution while the word2vec values exhibit a long tail of values above 0, suggesting that representation shift can occur even with relatively small frequency changes (i.e. low $\Delta \tau_\chi(w)$).

\subsection{Clustering Word Trajectories} 
Following prior work on concept drift~\cite{Gama:14} and time series analysis~\cite{Liao:05}, we seek to split the data automatically into coherent trajectories such as incremental increase, decrease and flatline (i.e. no change). For that we conduct an exploratory analysis with unsupervised clustering to identify common trajectories among concept drift and representation shift. By splitting the data according to these different kinds of change, we can find shared aspects across categories, such as similar words in the ``increase'' category for both kinds of change.

We transform the representation $\Delta \tau_e(w)$ and usage $\Delta \tau_f(w)$, tf-idf $\Delta \tau_\chi(w)$ time series using the LOWESS smoothing technique~\cite{Cleveland:81}, which smooths the spikes between time steps 
and thus makes the time series more amenable to clustering. Next, we use spectral clustering~\cite{Ng:02} with cosine distance as the distance between series, separately on the frequency $\Delta \tau_f(w)$, tf-idf $\Delta \tau_\chi(w)$ and representation $\Delta \tau_e(w)$ time series for all words in the vocabulary. We compare the average trajectories of the resulting clustered time series to investigate similarities in the two kinds of change (e.g. similar ``increase'' trajectories).

\subsection{Predicting Representation Shift}
Lastly, we can frame representation shift as a prediction problem: can we use frequency-based measures to predict change in meaning? Contrasting with previous studies~\cite{Kulkarni:15}, we look to predict the real value of change based on prior patterns of change as follows:
\begin{enumerate}
\item Representation shift $\Delta \tau_e(w) = \phi (\Delta \tau_e(w))$
\item Concept drift $\Delta \tau_e(w) = \phi(\Delta \tau_\chi(w))$
\item Concept drift and representation shift \\ $\Delta \tau_e(w) = \phi (\Delta \tau_\chi(w), \Delta \tau_e(w))$
\end{enumerate}

The second and third prediction tasks are especially interesting as they attempt to predict semantics using frequency data, which is novel to our knowledge.

\begin{figure*}[t!]
\vspace{-0.1cm}
	\centering
	\includegraphics[width=0.3\textwidth]{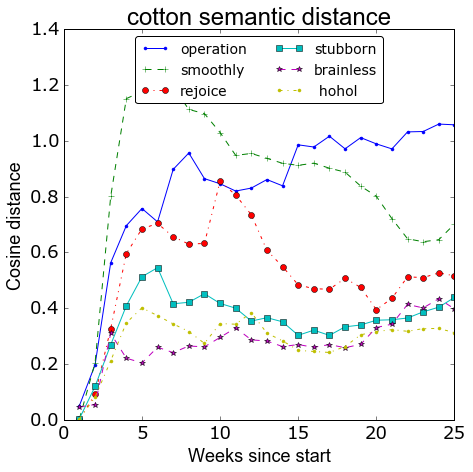}
	\includegraphics[width=0.3\textwidth]{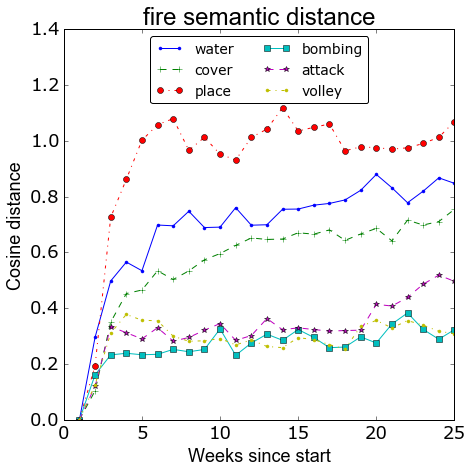}
	\includegraphics[width=0.3\textwidth]{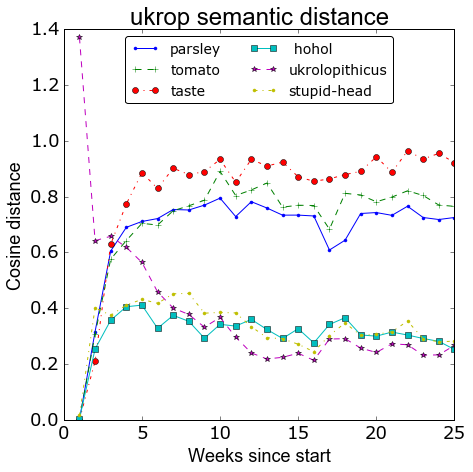}
	\includegraphics[width=0.32\textwidth]{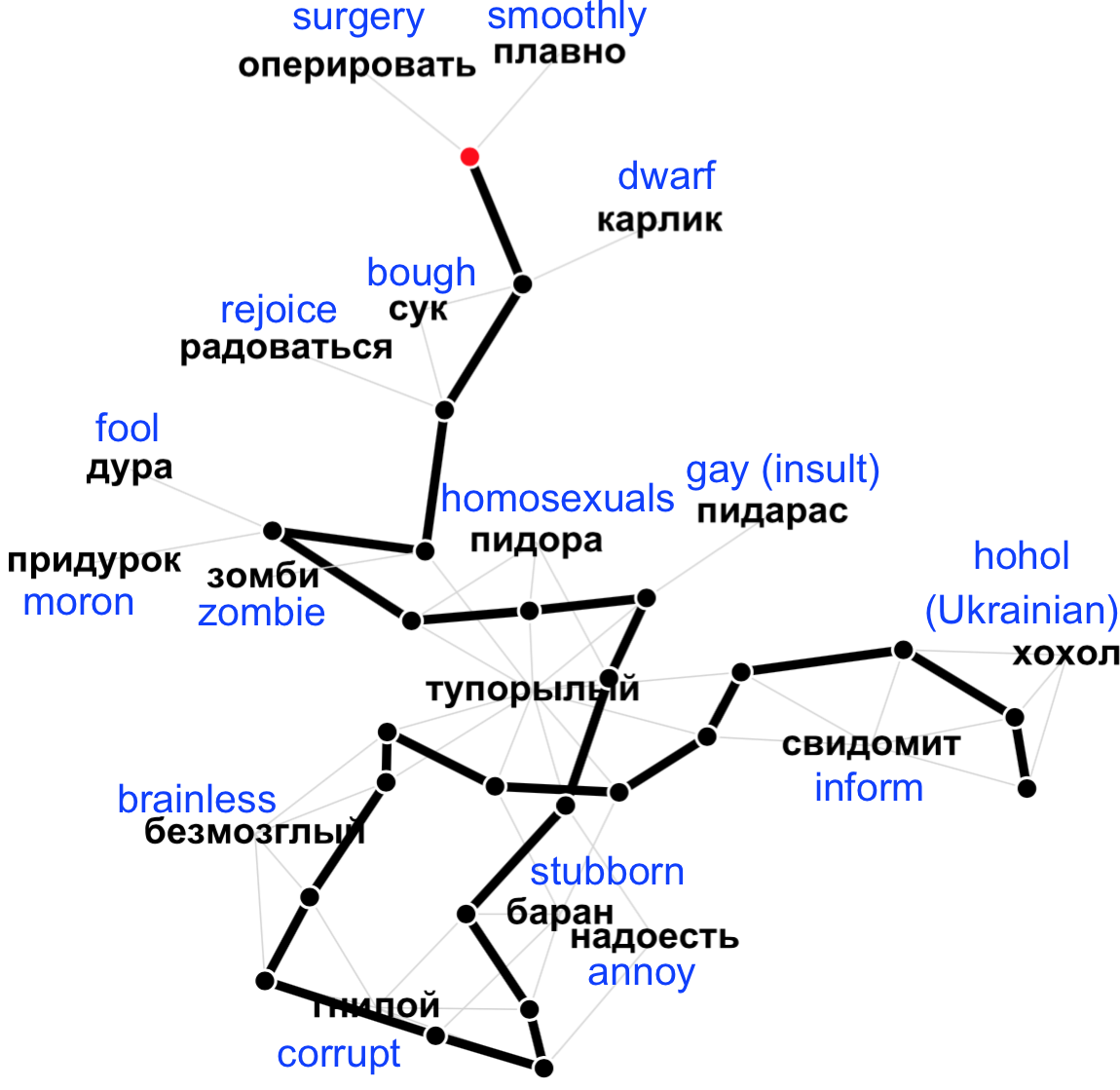}
	\includegraphics[width=0.32\textwidth]{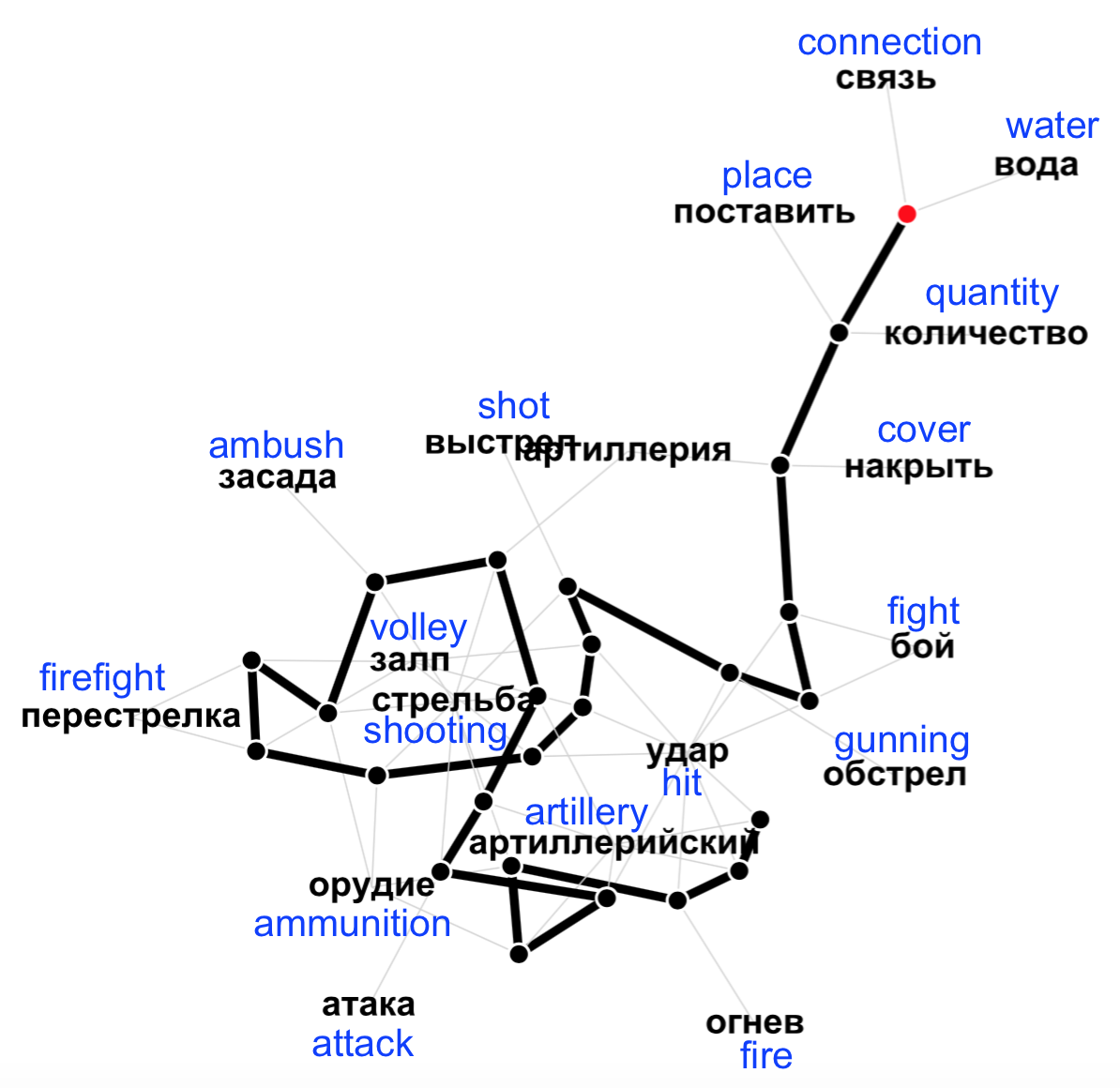}
	\includegraphics[width=0.34\textwidth]{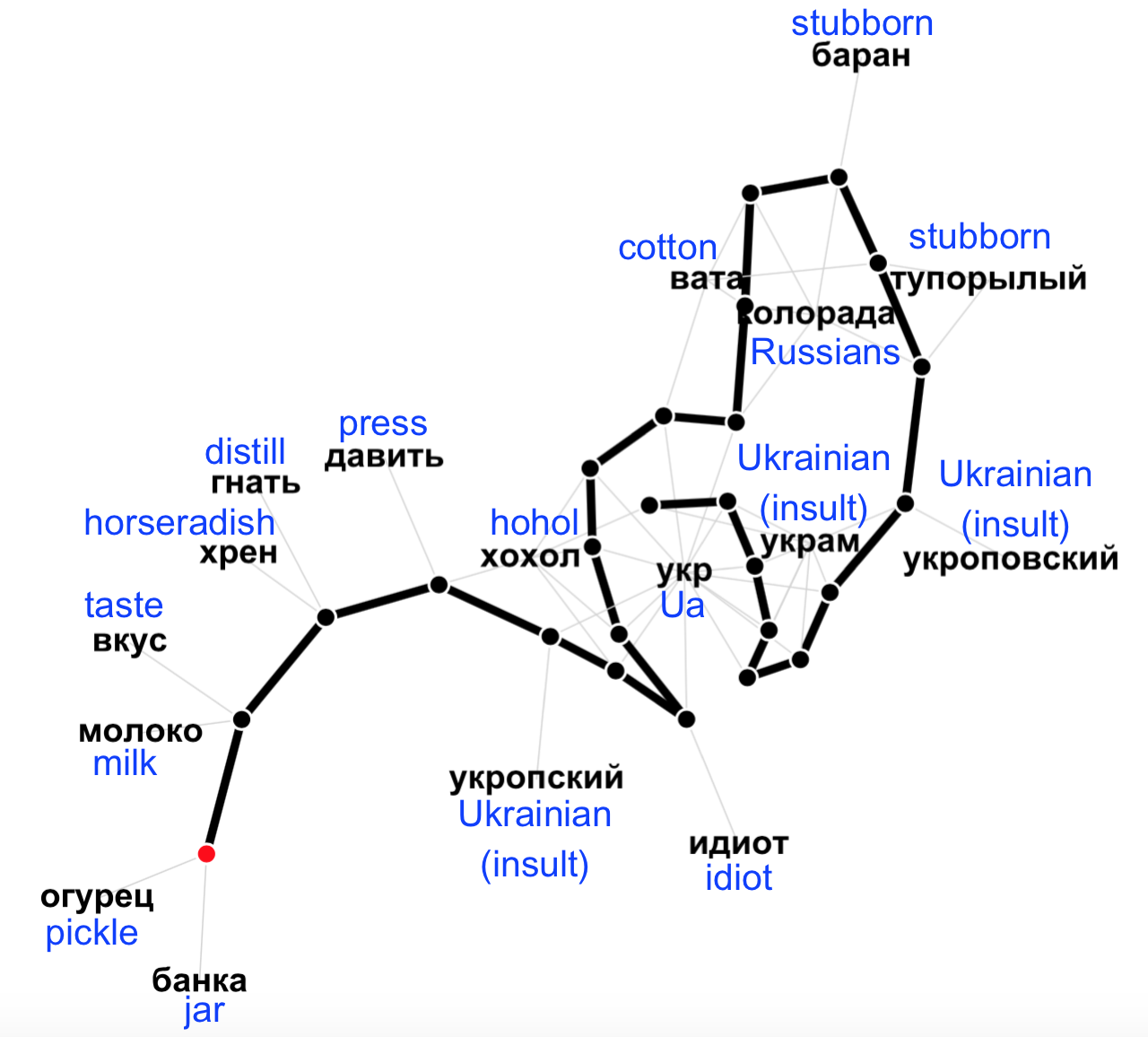}
	\caption{Representation shift $\Delta \tau_e(w)$ measured using cosine distance between each word's current representation and its original representation (top) and semantic trajectories of representation shift with two most similar words to the word of interest measure using Euclidian distance (bottom).}
	\label{fig:cos_traj}
\end{figure*}

For our experiment, we predict the final value in the time series by training with all data prior to the final time step. We experiment with forecasting and predict the value 1 to 3 weeks in advance using the full window. We perform this experiment using both representation shift $\Delta \tau_e(w)$ and concept drift $\Delta \tau_\chi(w)$ as the endogenous variables and representation shift $\Delta \tau_e(w)$ as the exogenous variable. This task gives us the most obvious test for whether representation shift is meaningful and detectable. We perform these experiments using 4-fold cross validation.

As a caveat, we restrict our prediction to words with consistent nonzero representation to avoid learning from overly sparse time series, i.e. remove words that have $\tau_{f}(w) = 0$ in over 50\% of the time steps. Although this cuts down our vocabulary to about 20,000 unique words, it also ensures that the estimators will not achieve artificially high accuracy by predicting zeros for the majority of the words. If we can predict the behavior of consistently dynamic words, we can justify the need to measure representation shift.

\paragraph{Model types}
Due to the noise in social media data, we avoided typical statistical methods in favor of more generalizable regression models. We implement a one-layer Long Short-Term Memory (LSTM) neural network\footnote{Theano via keras: https://keras.io/} for regression rather than classification~\cite{Oancea:14}. We initialize the network with one input node per timestep and a single output node, using the raw scalar output as the predicted value for regression. While relatively simple (e.g. single rather than multilayer), the network permits relatively short training time and thus a scalable framework for even larger datasets. We contrast the LSTM's performance with an AdaBoost regressor,\footnote{Scikit-learn: http://scikit-learn.org/stable/modules/\\generated/sklearn.ensemble.AdaBoostRegressor.html} since it represents a high-performing regression model comprising a collection of Random Forest regressors. We tested several other regression models, including SVMs with Linear and RBF kernels, with significantly worse performance and excluded them from our analysis. While we recognize that prediction of a word's representation shift is less informative than prediction of the word's actual representation (i.e. the embedding vector), we emphasize that predicting unexpected shifts, such as a particularly large shift, presents a useful application for our framework.

To provide a benchmark for these more complicated models, we rely on a simple baseline model that predicts the observed value of representation shift $\Delta \tau_e(w)$ at time $t$ using the value at time $t-1$.


\paragraph{Evaluation Metrics}

To test the models' performance, we report several standard evaluation metrics. We use the following notation: for each word $w$, $y_i$ is the observed value of representation shift $\Delta \tau_e(w)$ at time $t$, $\hat{y}_i$ is the predicted value at time $t$, $\bar y$ and $\bar{\hat{y}}$ denote the mean values over all words in the vocabulary.

We first report Pearson correlation $r$ as a measure of the linear dependence between the predicted and true values:
\begin{equation}
r = \frac{\sum_{i=1}^n (y_i -\bar{y}) (\hat{y}_i - \bar{\hat{y}}))}{\sqrt{\sum_{i=1}^n (y_i - \bar{y})^2}\sqrt{\sum_{i=1}^n (\hat{y}_i - \bar{\hat{y}})^2}}.
\label{eq:pearson}
\end{equation}

We also use Root Mean Squared Error $\gamma$ as a measure of the difference between predicted and true values:
\begin{equation}
\gamma = \sqrt{\frac{1}{n} \sum_{i=1}^n (y_i - \hat{y}_i)^2} \times 10^{-2}.
\label{eq:rmse}
\end{equation}



\subsection{Visualizing Meaning of New Words} 
We rely on visualization of short-term representation shift in order to demonstrate the practical application of our approach that can be used by analysts to investigate the meaning of new words emerging during a crisis {\it e.g., Titushky = street hooligans/government-paid thugs, DNR = Donetsk People's Republic, LNR = Luhansk People's Republic}~\cite{Katchanovski:14}. This kind of discovery would help to understand quickly the meaning of a new word based on words with a similar context. We use Principal Components Analysis (PCA) to project the 30-dimensional embeddings of each word $w_{t}$ at time $t$ to two dimensions, and for each time $t$ we also project the embedding for the two nearest neighbors of $w_{t}$. For example, the representation shift shown in Figure \ref{fig:motivation_trajectory} shows the movement of the 2-D projected {\it war} embedding over time toward more negative neighbor words like {\it bloodshed}.

\section{Results}

We present our results first with example visualizations of representation shift in existing words, the cluster analysis to compare concept drift and representation shift, the results of the representation shift predictive task, and lastly visualizations of representation shift in new words.

\begin{figure*}[t!]
\begin{subfigure}{.33\textwidth}
    \centering
    \includegraphics[width=0.75\linewidth]{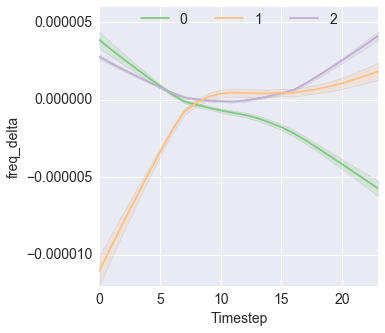}
    \caption{ $\Delta \tau_f(w)$ }
        \label{fig:freq_clusters}
\end{subfigure}
\begin{subfigure}{.33\textwidth}
    \centering
    \includegraphics[width=0.7\linewidth]{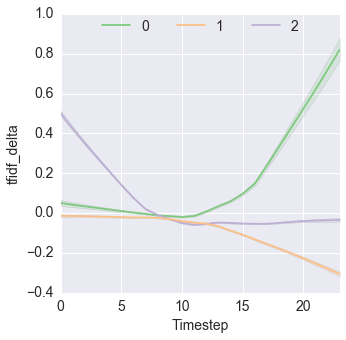}
    \caption{$\Delta \tau_\chi(w)$}
        \label{fig:tfidf_clusters}
\end{subfigure}
\begin{subfigure}{.33\textwidth}
    \centering
    \includegraphics[width=0.7\linewidth]{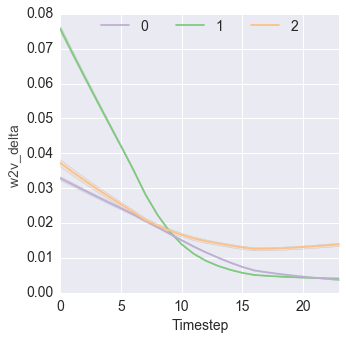}
    \caption{$\Delta \tau_e(w)$ }
        \label{fig:w2v_clusters}
\end{subfigure}
\caption{Clusters of word usage $\Delta \tau_f(w)$, $\Delta \tau_\chi(w)$ and representation shift $\Delta \tau_e(w)$  over time.}
\label{fig:dynamics_meanings_clusters}
\end{figure*}

\begin{table*}[t!]
\small
\centering
\begin{tabular}{ c  l  c  p{10cm} }
\sc Cluster & \sc Drift type & \sc Words & \sc Sample of most frequent words in each cluster  \\ \toprule[1.0pt]
0 & Increase & 21\% & soldier, Donetsk, hitman, ukrop (Ukrainian patriot), whizzbang, bogdan  \\ 
1 & Decrease & 34\% &  weapon death, protector, anniversary, mortar (weapon), rebels \\ 
2 & Flatline & 45\% & capitulation, Kharkov, Debaltseve, Ukrainian, rebels, hryvnia, product \\ 
\end{tabular}
\caption{Cluster details for $\Delta \tau_\chi(w)$ dynamics (Figure \ref{fig:tfidf_clusters}).}
\label{tab:dynamics_cluster_stats}
\end{table*}

\begin{table*}[t!]
\small
\centering
\begin{tabular}{ c  l  c  p{10cm} }
\sc Cluster & \sc Drift type & \sc Words & \sc Sample of most frequent  words in each cluster \\ \toprule[1.0pt]
0 & Flatline & 62\% & hryvnia, young, Bogdan, pepper, closed, suspect, sugar, Markin \\ 
1 & Flatline & 24\% & anniversary, cook, stage, Rus', die, mission, execute, girl, client\\ 
2 & Increase & 14\% & weapon, capitulation, soldier, Donetsk, Kharkov, battle of Debaltseve \\
\end{tabular}
\caption{Cluster details for representation $\Delta \tau_e(w)$ dynamics (Figure \ref{fig:w2v_clusters})}
\label{tab:meanings_cluster_stats}
\vspace{-0.2cm}
\end{table*}

\subsection{Visualizing Representation Shift}
To explore the shape of representation shift, for each word we can visualize not just the semantic distance from the previous timestep but also the distance since the beginning of time. We show the trajectory of the keywords {\it cotton, fire} and {\it dill} through representation space, in two different ways, in Figure~\ref{fig:cos_traj}. The top images show the distance from each keyword to six of its nearest neighbors: three from the beginning of the data and three from the end of the data. To contrast, the bottom images show the movement of the keyword through a 2-D projection of the embedding space, as well as the keyword's relative distance to its two nearest neighbors at each week, starting from the red point. 

We see that some words, such as {\it fire}, diverge quite cleanly from their original meaning in both the semantic distance over time of its neighbors (top figure) and in the stabilization toward the end of its movement to settle near words like {\it strike} (bottom). In the context of our data, this picture makes clear that the dominant context of {\it fire} is related to military words and that this tendency remains steady over time. The word {\it cotton} reveals a more surprising story: the shift away from concrete, medical words like {\it surgery} toward more subjective slang like {\it hohol} (pejorative for {\it Ukrainian}) reveals a trend of polarization that would warrant further content analysis from an analyst. As stated earlier, such a subtle shift in meaning would go undetected with frequency alone. 

\subsection{Clustering Word Dynamics}
We next explore our data with clustering to look for different general drift trends in both concept drift and representation shift, such as gradual increase, decrease and flatline~\cite{Gama:14}. We choose $c=3$ clusters with spectral clustering on a subsample of the vocabulary, using cosine distance between time series for the affinity in the clustering algorithm. We present an example of concept drift --  $\Delta \tau_\chi(w)$ and $\Delta \tau_f(w)$ and representation shift -- $\Delta \tau_e(w)$  in Figure \ref{fig:dynamics_meanings_clusters}.

The graphs show first-order differences rather than the raw statistics, and the $\Delta \tau_\chi(w)$ dynamics show an obvious split between increasing, decreasing and flatlining trends (positive, negative and near zero), while the $\Delta \tau_e(w)$ dynamics appear to universally trend toward either flatline or slight increase (near zero and above zero). The sudden drop at the beginning of the representation shift curves is the result of initial instability of the word embeddings, which is quickly corrected early in the time series. It is worth noting that neither statistic displayed bursty behavior in their clusters as might be expected from certain ``hot-topic'' words, which demonstrates the success of LOWESS smoothing.

We can further interpret these clusters by examining the distribution of words across them, shown by the statistics in Tables \ref{tab:dynamics_cluster_stats} and \ref{tab:meanings_cluster_stats}. The main items of interest are $\Delta \tau_{\chi}(w)$ cluster (2) in Table \ref{tab:dynamics_cluster_stats} and $\Delta \tau_{e}(w)$ cluster (0) in Table \ref{tab:meanings_cluster_stats}, which both show a trend toward increasing and feature similar military words such as ``soldier'' and ``Donetsk.'' Such a correspondence suggests analytic utility: an analyst provided with these clusters would be notified of the similar words undergoing concept drift and representation shift, and then given the option to break up the cluster into the individual word time series to further investigate the potential causes of such shift (e.g. a burst of ``soldier'' related news). We thus demonstrate how time series clustering can draw interesting parallels between concept drift and representation shift.


\subsection{Predicting Representation Shift}
In addition to comparing representation shift and concept drift, we explore the possibility of predicting representation shift using past shift and drift data. We present the results of prediction for representation shift from previous representation shift in Table~\ref{tab:predictions_shift_shift}.  Comparing models'  performance, we see clearly that the LSTM outperforms the AdaBoost model which outperforms the baseline, in both metrics. This shows that the LSTM picks up extra temporal signal that the other regressor missed, due to the LSTM's adaptive memory capability~\cite{Oancea:14}. 
Furthermore, we see that predicting one week ahead clearly surpasses forecasting for two or more weeks in all metrics and that the performance drops only slightly as we increase the distance of forecasting. This suggests that the signal for representation shift immediately before the period of prediction is nontrivial, reinforcing the conclusion that meaning shift can occur in a short timeframe. 

Next, we show the results of prediction for representation shift from previous concept drift in Table~\ref{tab:predictions_shift_drift}. We see an immediate decrease in performance as compared with the previous task as measured by the Pearson correlation coefficient, demonstrating the lack of signal associated. However, we note that the RMSE increased only slightly for both AdaBoost and LSTM as compared with the previous prediction task, suggesting that concept drift can provide nontrivial signal for representation shift. Similar to the previous task, the decrease in performance from forecasting one week to three weeks supports the short-term relationship between representation shift and concept drift: missing even a single week of concept drift data results in a sharp drop in performance.

To combine our signals, we predict representation shift as a function of both concept drift and representation shift, by concatenating the time series from $\Delta \tau_\chi(w)$ and $\Delta \tau_e(w)$ between timesteps 0 and $t-n$ to predict $\Delta \tau_e(w)$ at timestep $t$ (for forecasting n weeks). The results in Table \ref{tab:combined_prediction} show that this combined prediction performs somewhere between the other predictions, e.g. Pearson's correlation for combined prediction greater than $\Delta \tau_e(w) = \phi (\Delta \tau_\chi(w))$ prediction but less than $\Delta \tau_e(w) = \phi(\Delta \tau_e(w))$ prediction. The performance for 2-3 week prediction indicates that concept drift does contribute some signal to amplify the signal from representation shift, but the 1 week prediction results show lower performance due to concept drift noise. Note that the RMSE is comparable to the first prediction task (and lower for the LSTM), and thus the combined prediction has a competitive margin of error even if it does not produce a strong correlation. Overall, these models show robust performances relative to the baseline that would amplify an analyst's ability to preempt representation shift even within several weeks.

\begin{table}[t!]
\small
\begin{center}
\begin{tabular}{ l | c  c | c  c | c  c } 
& \multicolumn{2}{{c |}}{1 week} & \multicolumn{2}{{c |}}{2 weeks} & \multicolumn{2}{{c }}{3 weeks} \\ 
 & $r$ & $\gamma $ & $r$ & $\gamma $ & $r$ & $\gamma $ \\ \midrule[1.0pt]
Baseline & 0.62 & 5.00 & 0.30 & 5.59 & 0.29 & 4.83\\ 
AdaBoost & 0.69 & 5.73 & 0.40 & 6.56 & 0.39 & 5.97\\ 
 LSTM &  0.73 & 4.16 &  0.50 & 3.89 & 0.49 & 3.91\\ 
\end{tabular}
\end{center}
\caption{Prediction results for representation shift from previous representation shift: $\Delta \tau_e(w) = \phi (\Delta \tau_e(w))$. \\}
\label{tab:predictions_shift_shift}
\vspace{-0.4cm}
\end{table}

\begin{table}[t!]
\small
\begin{center}
\begin{tabular}{ l | c  c | c  c | c  c } 
& \multicolumn{2}{{c |}}{1 week} & \multicolumn{2}{{c |}}{2 weeks} & \multicolumn{2}{{c }}{3 weeks} \\
 & $r$ & $\gamma $ & $r$ & $\gamma $ & $r$ & $\gamma $ \\ \midrule[1.0pt]
Baseline & 0.17 & 47.4 & 0.28 & 59.5 & 0.29 & 43.3\\ 
AdaBoost & 0.40 & 7.26 & 0.16 & 7.89 & 0.15 & 8.30\\ 
LSTM & 0.44 & 5.56 & 0.21 & 4.37 & 0.18 & 4.39\\ 
\end{tabular}
\end{center}
\caption{Prediction results for representation shift from previous concept drift: $\Delta \tau_e(w) = \phi (\Delta \tau_\chi(w))$.}
\label{tab:predictions_shift_drift}
\vspace{-0.2cm}
\end{table}

\begin{table}[t!]
\small
\begin{center}
\begin{tabular}{ l | c  c | c  c | c  c }
& \multicolumn{2}{{c |}}{1 week} & \multicolumn{2}{{c |}}{2 weeks} & \multicolumn{2}{{c }}{3 weeks} \\ 
 & $r$ & $\gamma $ & $r$ & $\gamma $ & $r$ & $\gamma $ \\ \midrule[1.0pt]
Baseline & 0.35 & 5.19 & 0.30 & 5.37 & 0.21 & 4.34 \\
AdaBoost & 0.47 & 5.24 & 0.49 & 6.43 & 0.40 & 4.46 \\ 
LSTM & 0.52 & 3.21 & 0.52 & 4.29 & 0.48 & 2.90 \\ 
\end{tabular}
\end{center}
\caption{Prediction results for representation shift from representation shift and concept drift: \\ $\Delta \tau_e(w) = \phi (\Delta \tau_\chi(w), \Delta \tau_e(w))$. }
\label{tab:combined_prediction}
\vspace{-0.3cm}
\end{table}

\begin{figure*}[t!]
	\centering
	\begin{subfigure}[b]{0.32\textwidth}
	\includegraphics[width=\textwidth]{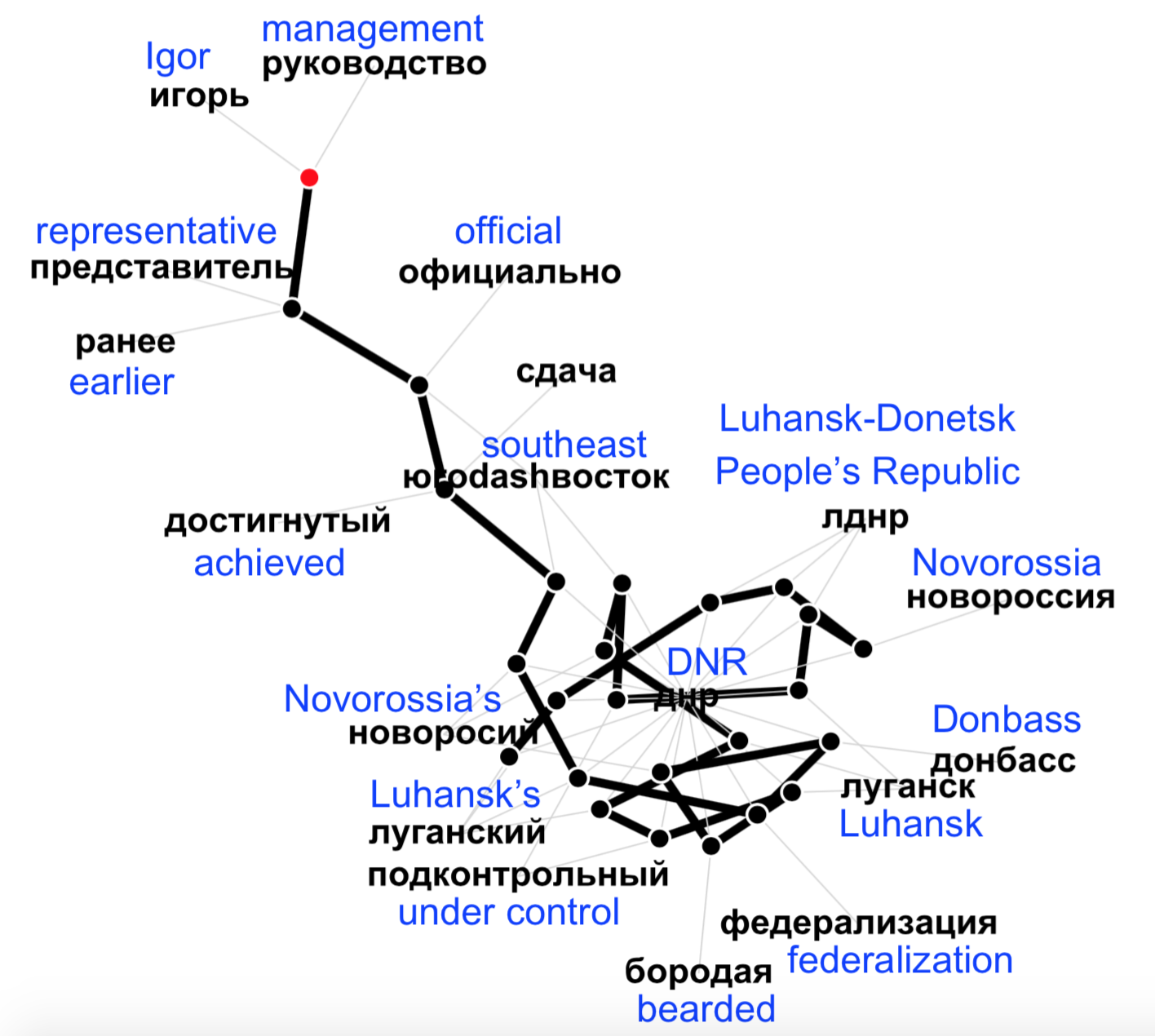}
	\caption{LNR: Luhansk People's Republic}
	\label{fig:lnr_shift}
	\end{subfigure}
	\begin{subfigure}[b]{0.32\textwidth}
	\includegraphics[width=\textwidth]{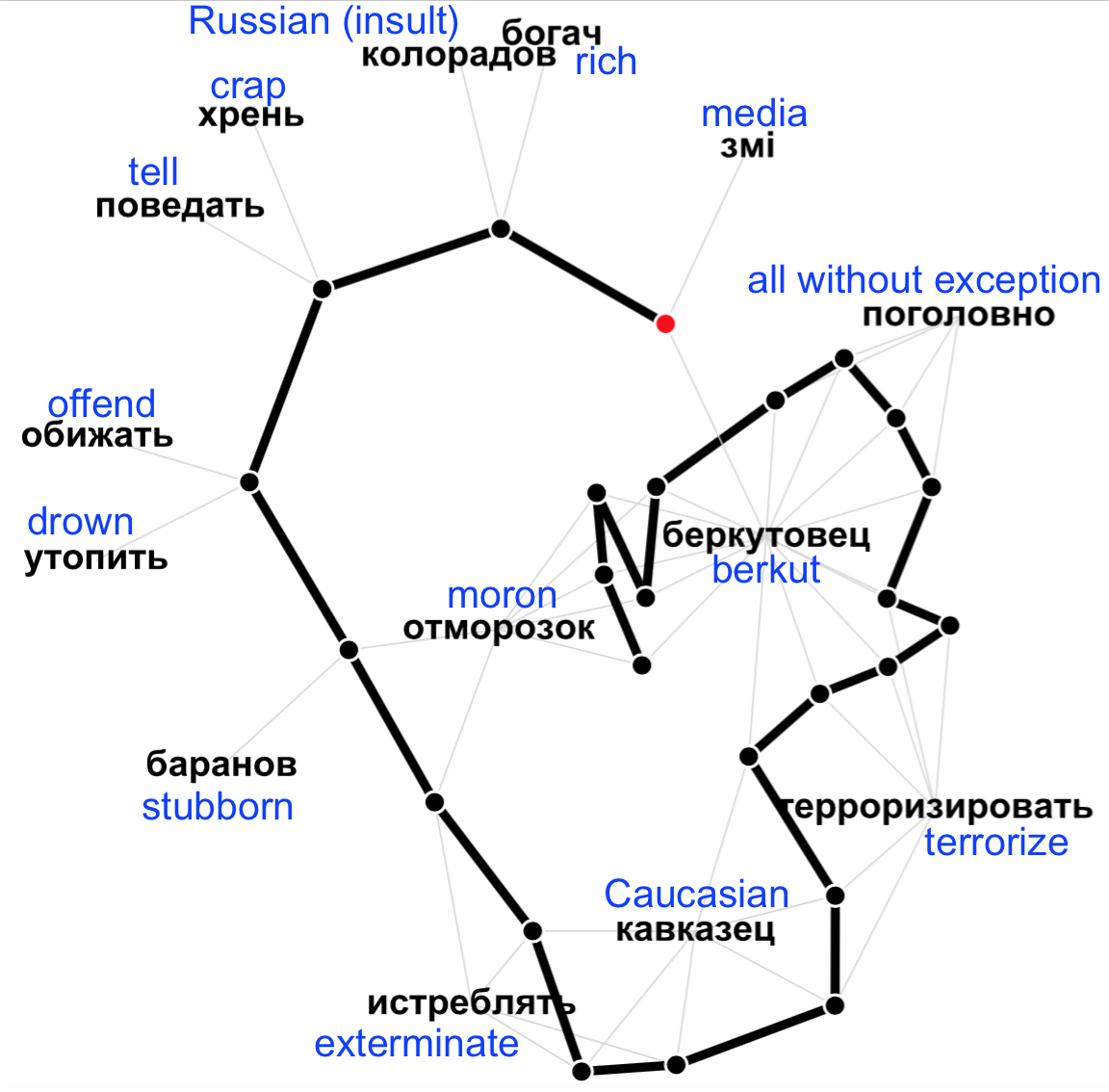}
	\caption{Titushky}
	\label{fig:titushky_shift}
	\end{subfigure}
	\begin{subfigure}[b]{0.32\textwidth}
	\includegraphics[width=\textwidth]{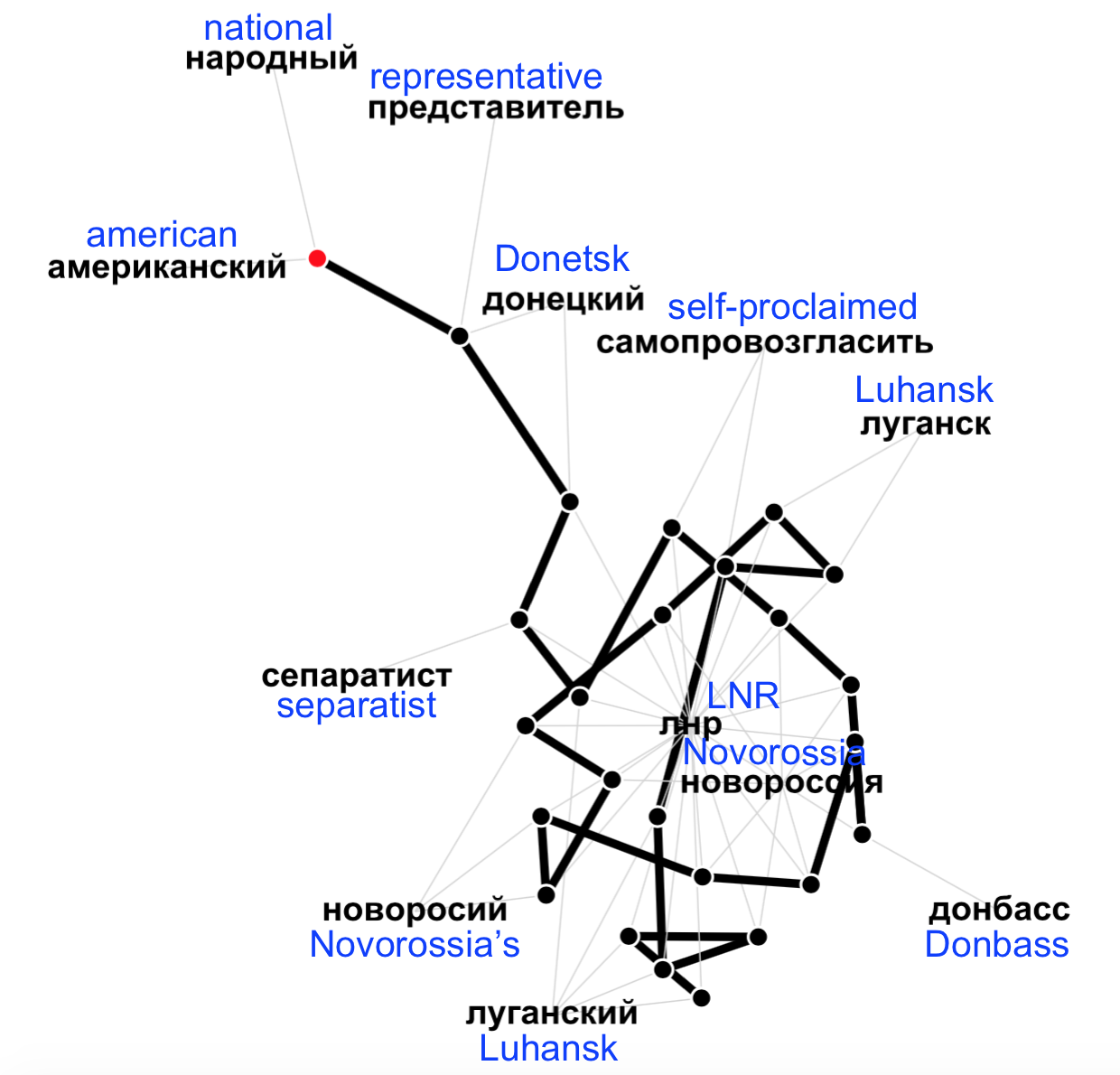}
	\caption{DNR: Donetsk People's Republic}
	\label{fig:dnr_shift}
	\end{subfigure}
	\caption{Semantic trajectories of newly emerging terms during crisis.}
	\label{fig:new_words}
\end{figure*}

To investigate the source of error in our predictions, we look at the performance of the trained LSTM on a set of keywords related to the Ukraine-Russia conflict. These words are shown in Table \ref{fig:error_examples} with the relative errors $\epsilon_{rel} = | \frac{y_i - {\hat{y}_i}}{y_i} |$ in their representation shift, generated from forecasting 1 week to predict $\Delta \tau_e(w)$ from $\Delta \tau_\chi(w)$. We see that common nouns (e.g. {\it war}) tend toward lower error while keywords with higher errors are more often proper nouns (e.g. {\it Donetsk}). This suggests that representation shift is more predictable in common nouns than in proper nouns, perhaps due to exogenous influences such as real-world events that influence representation shift more than frequency alone~\cite{Kulkarni:15}. Thus, the prediction framework may best serve analysts as a way to preempt the representation shift in common nouns such as ``help'' which would otherwise go unnoticed with frequency alone.

While room for improvement remains, our tests demonstrate that short-term representation shift can be accurately predicted with models such as an LSTM.

\begin{table}[t!]
\small
\centering
\begin{tabular}{ l c | l c }
Keyword & Error & Keyword & Error \\ \toprule[1.0pt]
LNR & 0.895 & Crimea & 0.595 \\ 
militiaman & 0.818 & Russia & 0.557 \\ 
Donetsk & 0.782 & Poroshenko & 0.546 \\ 
Putin & 0.772 & USA & 0.519 \\ 
Novorussia & 0.700 & Kiev & 0.464 \\ 
DNR & 0.677 & ukrop & 0.462 \\  
Moscow & 0.667 & help & 0.368 \\ 
ATO & 0.650 & Ukraine & 0.265 \\ 
fire & 0.632 & junta & 0.231 \\ 
negotiations & 0.628 & war & 0.015 \\ 
\end{tabular}
\caption{Relative errors on keywords, for forecasting representation shift from concept drift (1 week in advance).}
\label{fig:error_examples}
\vspace{-0.3cm}
\end{table}


\subsection{Discovering Meaning of Emerging Terms}

Moving on from prediction, we now show how visualizing representation shift can be used to uncover the meaning of new words. In Figure \ref{fig:new_words}, we present a 2-D projection of representation shift for three words that emerged during the crisis: {\it LNR} (Luhansk People's Republic), {\it Titushky} (mercenaries), and {\it DNR} (Donetsk People's Republic). We see that these new words do not have identical trajectories in their representation shift. For instance, {\it Titushky} has a more gradual transition to its final negative definition (e.g., {\it moron}) while {\it DNR} moves quickly toward other words related to locations relevant to the crisis (e.g., {\it Luhansk}).

In addition to displaying the shift trajectory, this visualization can help the analyst to discover the meaning of previously unobserved words based on their nearest neighbors. In the dynamic real-time nature of social media, a semantic representation of a new lexical item provides the ability for standard keyword or topically based searches to be modified in real-time to include or remove terms. Instead of a traditional fixed query, our method allows for exploration of content to naturally follow semantic changes in terms (e.g. switching between the separate shifts in {\it Titushky} and {\it moron}) - key to capturing relevant content in the rate-limited API world of social data.

Such a visual representation could replace the need for a dictionary: rather than attempting to parse or translate the term on its own, the analyst would merely need to check its representation shift to determine the relevant context for the term. This would also help the analyst determine the best points in time to study the text of posts containing the word in question, such as how people were discussing the word {\it DNR} in the most early versus late stage of the Ukraine crisis. Lastly, a system generating such visualizations could recommend interesting words for the analyst to explore further, based on how surprising their change trajectory appears. This is especially critical given the volume of new or reoccurring words that emerge on a daily basis in social media.

Using several new Russian words as examples, we argue that tracking the representation shift of new words can help stay up to date with sudden language change.


\subsection{Discussion}

Our exploratory analysis of representation shift has revealed its utility in picking up unexpected trends, such as the shift in common words like {\it cotton}. Both the visualizations and the prediction task outlined above could serve an analyst well in another crisis situation similar to the Ukraine-Russia conflict. When words like {\it Titushky} are rapidly introduced to the lexicon or when existing words like {\it cotton} undergo sudden polarization, a system that can visualize and preempt such shifts will help an analyst stay on top of the crisis.

Our work also has implications for downstream applications like event detection, summarization, and detection of misinformation. For example, a word whose frequency increases suddenly but whose representation remains static could be the result of a misinformation campaign by spam-bots\footnote{http://www.bbc.com/news/technology-16108876} using the word repeatedly without changing its context. An effect shift-detection system will present semantic information to analysts in a transparent and actionable way, e.g., highlighting words that have a high likelihood of shift in the near future. To that end, future work will need to determine what kinds of representation shift (e.g. gradual versus sudden) present the most useful insight to analysts and how to present representation shift with a system that is interactive and informative. 

In the future, we hope to extend our approach to a cross-country analysis that compares location-specific patterns of representation shift. This would allow analysts to further explore why a certain country or region responded to an event differently than the others, e.g., if {\it fire} became more associated with {\it war} in Russia versus Ukraine. Furthermore, combining distributional semantics with sentiment
could reveal how certain words such as {\it Putin} can become semantically polarized over time, with one country expressing an unusually positive view of the word while another country uses the word more negatively. 

Our analysis's main limitation was its focus on stemmed word unigrams rather than bigrams or other linguistic units with a larger context. For instance, polysemous words such as ``bank'' may have a less coherent representation than words with a single, concrete meaning~\cite{Trask:15}. We may also need to test metrics other than cosine distance between a word's vectors to measure representation shift over time, such as the overlap between a word's k-nearest neighbors at each timestep~\cite{Hamilton:16Cultural}.
In addition, we do not compare the representation shifts of our data with well-understood shifts in previous experiments such as \citeAuthorYear{Kulkarni:15}, whose data has not been made public. We are interested less in comparing prior results on representation shift and more in testing several methods to characterize representation shift in social media, rather than more formal corpora like Google Books. 
 Lastly, our data's range of 25 weeks may be too short to cover a meaningful shift in distributional semantics, as prior work suggests that lasting linguistic change occurs over the course of decades or even centuries~\cite{Hamilton:16Diachronic,Sagi:09}. Despite the dynamic nature of social media, we may need to expand our timeframe from weeks to years, to reliably tie representation shift to concept drift. Nonetheless, our work suggests that detecting short-term shift, such as the new meaning of the word {\it ukrop}, can highlight changes that may have gone unnoticed.
 
\section{Conclusion}
This work provides a generalizable proof of concept for future studies on short-term representation shift in social media -- despite noisy data, the word vector representations generated are robust.  Our prediction results show that by considering representation in addition to raw frequency, we are able not only to forecast meaning change for words over time but also to isolate interesting words, i.e. those with dynamic contexts. We propose representation shift as a novel metric to track unexpected changes during a crisis, showing the power of semantics in action.

\bibliographystyle{aaai}
\bibliography{references}

\section{Appendix}

\paragraph{Predicting Country-Specific Representation Shift} In addition to making representation shift predictions over the full data, we also provide predictions over country-specific subsets of the data. We expected that subsampling the data in this way would provide more accurate predictions, since location-specific models would partly control for geographic variation. However, we find that separating the data by country makes prediction more difficult, and Table \ref{tab:ru_ua_prediction} shows the evaluation metrics for both Russia- and Ukraine-specific predictions, which clearly demonstrate a reduced performance.

\begin{table}[htb!]
\small
\vspace{-0.25cm}
\begin{center}
\begin{tabular}{| l | c | c | c | c  |} \hline
Model & Pearson & $\chi \times 10^{-2}$ & $\zeta \times 10^{2}$  \\ \toprule[2.0pt]
\multicolumn{4}{{|c|}}{Forecasting (1 week) (RU)} \\ \hline
Baseline & 0.08 & 2.52 & 400 \\ \hline
AdaBoost & 0.10 & 2.22 & 143 \\ \hline
\rowcolor[gray]{0.8} LSTM & 0.19 & 1.92 & 108 \\ \hline
\multicolumn{4}{{|c|}}{Forecasting (2 weeks) (RU)} \\ \hline
Baseline & 0.03 & 3.61 & 194 \\ \hline
AdaBoost & 0.09 & 2.78  & 124 \\ \hline
\rowcolor[gray]{0.8} LSTM & 0.19 & 1.92  & 91 \\ \hline
\multicolumn{4}{{|c|}}{Forecasting (3 weeks) (RU)} \\ \hline
Baseline & 0.05 & 2.30 &295 \\ \hline
AdaBoost & 0.09 & 2.68 & 219 \\ \hline
\rowcolor[gray]{0.8} LSTM & 0.18 & 1.93 &  89 \\ \hline
\multicolumn{4}{{|c|}}{Forecasting (1 week) (UA)} \\ \hline
Baseline & 0.16 & 4.97 & 160 \\ \hline
AdaBoost & 0.14 & 7.37 & 721 \\ \hline
\rowcolor[gray]{0.8} LSTM & 0.27 & 4.27 &  187 \\ \hline
\multicolumn{4}{{|c|}}{Forecasting (2 weeks) (UA)} \\ \hline
Baseline & 0.16 & 6.59 & 351 \\ \hline
AdaBoost & 0.12 & 7.29 &  769 \\ \hline
\rowcolor[gray]{0.8} LSTM & 0.24 & 4.31 & 179 \\ \hline
\multicolumn{4}{{|c|}}{Forecasting (3 weeks) (UA)} \\ \hline
Baseline & 0.10 & 5.17  & 202 \\ \hline
AdaBoost & 0.11 & 7.41 & 635 \\ \hline
\rowcolor[gray]{0.8} LSTM & 0.22 & 4.33 & 182 \\ \hline
\end{tabular}
\end{center}
\caption{Prediction results for representation shift from previous representation shift $\Delta \tau_e(w) = \phi (\Delta \tau_e(w))$ for Russia (RU) and Ukraine (UA).}
\label{tab:ru_ua_prediction}
\vspace{-0.3cm}
\end{table}

In addition to Pearson correlation and Root Mean Squared Error $\gamma$, we also report Maximum Absolute Percent Error (MAPE), which is a measure of the magnitude of the maximum percent difference between predicted and true values:
$$\zeta = \left ( \max  \frac{| y_i - \hat{y}_i |}{y_i} \right ) \times 100.
$$

\begin{figure}[htb!]
	\centering
	\includegraphics[width=0.4\textwidth]{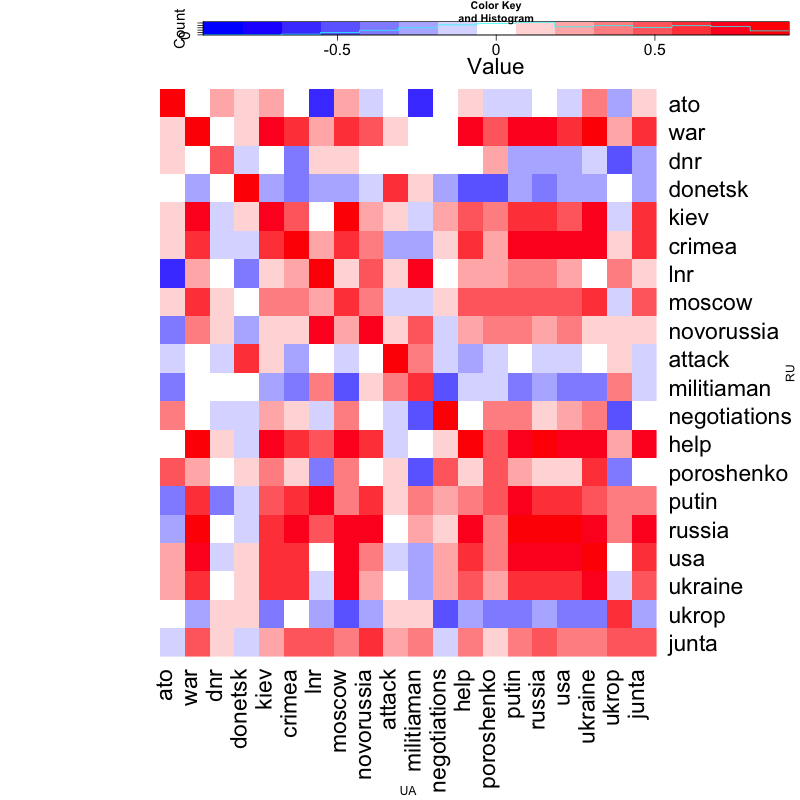}
	\caption{Pearson correlation between keyword dynamics $\Delta \tau_\chi(w)$  in Russia and Ukraine.}
	\label{fig:corr_ru_ua}
\end{figure}

While the predictors still perform comparatively as expected (LSTM > AdaBoost > Baseline), the Pearson's coefficients fall well below the level expected based on the original representation shift prediction. We suspect that the majority of the errors result from  predicting change when no change actually occurred, as suggested by the MAPE ($\zeta$) values. Such a poor result likely results from the relative sparsity in the country-specific data, i.e. some of the vocabulary might occur only in one country as has also been discussed by~\cite{Gladkova:16}. 
\paragraph{Correlating Word Dynamics between Countries}
In Figure~\ref{fig:corr_ru_ua} we present Pearson correlation between keyword dynamics estimated using TFIDF score in Russian vs. Ukraine. We observe that word dynamics have positive correlation between two countries (Ru and Ua) for such keywords as Russia, Ukraine, Putin and war. In contrast, keywords like LNR (Luhansk People's Republic) and ATO (Anti-Terrorist Operation) are negatively correlated.  These results will have implications in spatiotemporal variations for trend detection methods, which has largely relied on frequency-based measures~\cite{Hendrikson:15}.

\begin{figure}[b!]
	\centering
	\begin{subfigure}[b]{0.4\textwidth}
	\includegraphics[width=\textwidth]{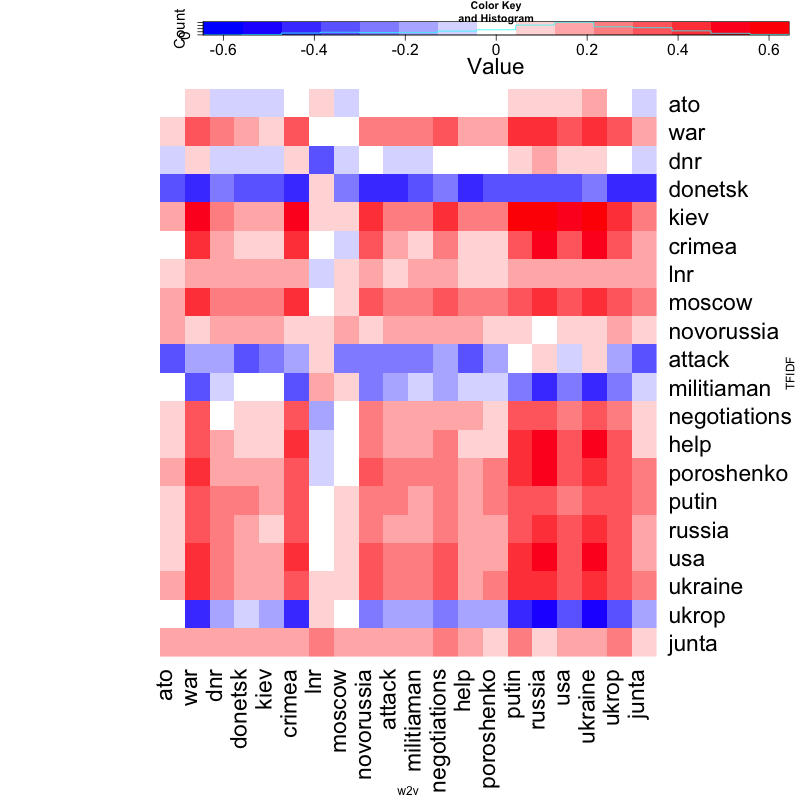}
	\caption{Russia (RU)}
	\label{fig:ru}
	\end{subfigure}
	\vspace{0.5cm}
	\begin{subfigure}[b]{0.4\textwidth}
	\includegraphics[width=\textwidth]{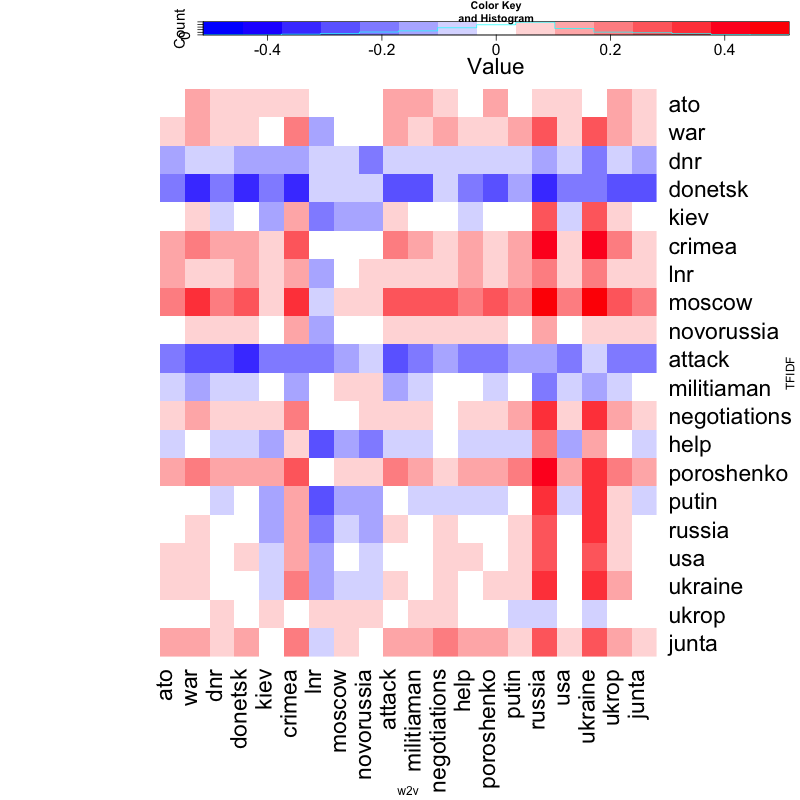}
	\caption{Ukraine (UA)}
	\label{fig:ua}
	\end{subfigure}
	\caption{The relationship between representation shift ($\Delta \tau_e(w)$ -- w2v) and word dynamics ($\Delta \tau_\chi(w)$ -- TFIDF) estimated using Pearson correlation in RU and UA.}
	\label{fig:corr_tfidf_w2v_ru_ua}
\end{figure}

\paragraph{Correlating Word Usage and Representation}
Figure~\ref{fig:corr_tfidf_w2v_ru_ua} demonstrates correlations between word dynamics $\Delta \tau_\chi(w)$ and representation shift  $\Delta \tau_e(w)$  in each country.
These findings allow to further analyze the degree to which words embeddings encode frequency information, and thus, likely to be biased by frequency dynamics as has been recently addressed by~\cite{Schnabel:15}.

\end{document}